\documentclass[draft,jgrga]{agutex2015}

 \usepackage{colortbl}
 \usepackage{multirow}
  \usepackage{graphicx}
  \usepackage{color}
  \usepackage{amssymb}
  \usepackage{caption}
  \usepackage{rotating}

\authorrunninghead{GENESTRETI ET AL.}

\titlerunninghead{OHM'S LAW FOR NEW MMS EDR EVENT}

\begin{document}

%
%

\title{MMS observation of asymmetric reconnection supported by 3-D electron pressure divergence}
%
%

%
%

\authors{K.~J. Genestreti,\altaffilmark{1}
	      A. Varsani, \altaffilmark{1} \thanks{now at Mullard Space Science Laboratory, University College London, Dorking, UK.} 
	      J.~L. Burch, \altaffilmark{2}
	      P.~A. Cassak, \altaffilmark{3}
	      R.~B. Torbert, \altaffilmark{4,2}
	      R. Nakamura, \altaffilmark{1}
	      R.~E. Ergun, \altaffilmark{5,6}
	      T.~D. Phan, \altaffilmark{7} 
	      S. Toledo-Redondo, \altaffilmark{8}
	      M. Hesse, \altaffilmark{9}
	      S. Wang, \altaffilmark{10}
 	      B.~L. Giles, \altaffilmark{11}
 	      C.~T. Russell, \altaffilmark{12} 
	      Z. V\"or\"os, \altaffilmark{1}
	      K.-J. Hwang, \altaffilmark{2}
	      J.~P. Eastwood, \altaffilmark{13}
	      B. Lavraud, \altaffilmark{14}
	      C.~P. Escoubet, \altaffilmark{15}
	      R.~C. Fear, \altaffilmark{16}
	      Y. Khotyaintsev, \altaffilmark{17}
	      T.~K.~M. Nakamura, \altaffilmark{1}
	      J.~M. Webster, \altaffilmark{18}
	      W. Baumjohann \altaffilmark{1}
}

\altaffiltext{1}{Space Research Institute, Austrian Academy of Sciences, Graz, Austria.}
\altaffiltext{2}{Southwest Research Institute, San Antonio, TX, USA.}
\altaffiltext{3}{West Virginia University, Morgantown, WV, USA.}
\altaffiltext{5}{Department of Astrophysical and Planetary Sciences, University of Colorado, Boulder, CO, USA.}
\altaffiltext{6}{Laboratory for Atmospheric and Space Physics, University of Colorado, Boulder, CO, USA.}
\altaffiltext{7}{Space Science Institute, University of California Berkeley, Berkeley, CA, USA.}
\altaffiltext{8}{Science Directorate, European Space Agency, ESAC, Madrid, Spain.}
\altaffiltext{9}{Birkeland Centre for Space Science, University of Bergen, Norway.}
\altaffiltext{10}{University of Maryland, College Park, MD, USA.}
\altaffiltext{11}{Heliophysics Science Division, NASA Goddard Space Flight Center, Greenbelt, MD, USA.}
\altaffiltext{12}{University of California, Los Angeles, CA, USA.}
\altaffiltext{13}{The Blackett Laboratory, Imperial College London, London, UK}
\altaffiltext{14}{Institute de Recherche en Astrophysique et Plan\'etologie, Toulouse, France.}
\altaffiltext{15}{Space Science Department, European Space Agency, Noordwijk, Netherlands.}
\altaffiltext{16}{Department of Physics \& Astronomy, University of Southampton, Southampton, UK.}
\altaffiltext{17}{Swedish Institute of Space Physics, Uppsala, Sweden.}
\altaffiltext{18}{Rice University, Houston, TX, USA.}









%
%


\keypoints{\item{We analyze MMS data measured during a slow crossing of the density-asymmetric magnetopause}
		 \item{Ion and electron dynamics are consistent with a normal crossing of an inner diffusion region}
		 \item{$\vec{J}\cdot\vec{E}'$ appeared to result from in and out-of-plane gradients of gyrotropic and agyrotropic electron pressure tensor}}


%
%


\begin{abstract}
We identify the electron diffusion region (EDR) of a guide-field dayside reconnection site encountered by the Magnetospheric Multiscale (MMS) mission and estimate the terms in generalized Ohm's law that controlled energy conversion near the X-point. MMS crossed the moderate-shear ($\sim$130$^\circ$) magnetopause southward of the exact X-point. MMS likely entered the magnetopause far from the X-point, outside the EDR, as the size of the reconnection layer was less than but comparable to the magnetosheath proton gyro-radius, and also as anisotropic gyrotropic ``outflow'' crescent electron distributions were observed. MMS then approached the X-point, where all four spacecraft simultaneously observed signatures of the EDR, e.g., an intense out-of-plane electron current, moderate electron agyrotropy, intense electron anisotropy, non-ideal electric fields, non-ideal energy conversion, etc. We find that the electric field associated with the non-ideal energy conversion is (a) well described by the sum of the electron inertial and pressure divergence terms in generalized Ohms law though (b) the pressure divergence term dominates the inertial term by roughly a factor of 5:1, (c) both the gyrotropic and agyrotropic pressure forces contribute to energy conversion at the X-point, and (d) both out-of-the-reconnection-plane gradients ($\partial/\partial M$) and in-plane ($\partial/\partial L,N$) in the pressure tensor contribute to energy conversion near the X-point. This indicates that this EDR had some electron-scale structure in the out-of-plane direction during the time when (and at the location where) the reconnection site was observed.
\end{abstract}

%
%

%

\begin{article}

%
%

\section{Introduction}

During its magnetopause survey phase, NASA's Magnetospheric Multiscale (MMS) mission encountered a large number of plasma-kinetic-scale magnetic reconnection sites \cite{Fuselier.2017,Wang.2017}. One of the key objectives of MMS is to investigate the kinetic processes that drive reconnection in the electron diffusion region (EDR) for a variety of upstream conditions \cite{Burch.2016a}. To this end, each magnetopause EDR observation contributes to the central goals of MMS, as it allows for events with similar and dissimilar conditions to be compared and contrasted.

\subsection{Energy conversion and electron dynamics during reconnection}

\cite{Burch.2016b} identified the role of agyrotropic ``crescent-shaped'' electron velocity distribution functions (eVDFs) in energy conversion in the central EDR of high-magnetic-shear reconnection, as was predicted by \cite{Hesse.2014}. Similarly, \cite{BurchandPhan.2016} found agyrotropic crescents in the central EDR of moderate-shear reconnection, which was similarly predicted by \cite{Hesse.2016}. \cite{Khotyaintsev.2016}, \cite{Phan.2016}, and \cite{Hwang.2017} found that downstream of the central EDR, gyrotropic and anisotropic ``outflow crescent'' eVDFs can be observed. Similar to the agyrotropic crescents of \cite{Burch.2016b} and \cite{Hesse.2014,Hesse.2016}, these outflow crescents are a signature of the mixing of inflowing plasmas between the reconnection X-point and the electron stagnation point \cite{Shay.2016}. 

\cite{Ergun.2016b,Ergun.2017} reported observations of large-amplitude parallel electrostatic waves, which they associated with drift-instability-driven ``corrugations'' of the magnetopause near the separatrix and X-point. \cite{Cassak.2017} and \cite{Genestreti.2017} reported $\vec{J}\cdot(\vec{E}+\vec{v}_e\times\vec{B})>0$, the local per-volume rate of work done on the plasma by the non-ideal electric field, in central EDRs that were several orders of magnitude larger than what was predicted by 2.5-d particle-in-cell (PIC) simulations. (Here, the current density vector is $\vec{J}$, the electric and magnetic fields are $\vec{E}$ and $\vec{B}$, the electron bulk velocity is $\vec{v}_e$, and the electron-convective-frame electric field is defined as $\vec{E}'\equiv\vec{E}+\vec{v}_e\times\vec{B}$). While the local $\vec{J}\cdot\vec{E}'$ can be much larger than predicted, it may not be indicative of a larger-than-predicted global reconnection rate \cite{Cassak.2017}.

\cite{Nakamura.2017}, \cite{Price.2016,Price.2017}, and \cite{Le.2017} have all recently performed 3-d PIC simulations of MMS EDR events. \cite{Nakamura.2017} analyzed a large-scale simulation of the very large guide field Kelvin-Helmholtz vortex reconnection event of \cite{Eriksson.2016}. Both the simulation and the data showed parallel out-of-plane electric fields in the $\vec{J}\cdot\vec{E}'>0$ region that were several times larger than what is expected for the nominal fast reconnection rate of 0.1. \textit{Nakamura et al.} suggested that these large-amplitude electric fields were a result of their reconnection driven by the vortex flow rather than being spontaneous. \cite{Price.2017} found that small-scale turbulence developed in their simulation, causing large-amplitude parallel electric fields and structure to form in the $M-N$ plane of the current layer (see Section 2 for $LMN$ coordinate definition). This structured magnetopause was similar in character to the corrugated magnetopause of \cite{Ergun.2017} and the lower hybrid drift turbulence of \cite{Roytershteyn.2012}. \citeauthor{Price.2017} noted that the wrapping of the normal electric field $E_N$ into the direction of the current ($M$) resulted in greatly enhanced $\vec{J}\cdot\vec{E}'$. \cite{Le.2017}, which analyzed a 3-d simulation of the same event studied by \cite{Price.2016,Price.2017}, found that these oscillations in the current layer caused intense parallel electron heating as was observed by MMS \cite{Burch.2016b}. \textit{Price et al.} noted that the turbulence caused significant anomalous resistivity in the dissipation region, but did not affect the formation of the agyrotropic crescent eVDFs predicted by laminar 2.5-d PIC simulations of reconnection \cite{Hesse.2014,Hesse.2016}. 

\subsection{$\vec{J}\cdot\vec{E}'$ and generalized Ohm's law}

\cite{Hesse.2014,Hesse.2016} investigated the sources of the reconnection electric field in the central diffusion region of asymmetric reconnection with 2.5-d PIC simulations. For simulations of anti-parallel and guide field ($B_M/B_L\sim1$) reconnection, \citeauthor{Hesse.2014} determined each of the terms in generalized Ohm's law, which is:

\begin{equation}
\vec{E}'=\eta\vec{J}-\frac{1}{en}\nabla\cdot\bar{P}_e+\frac{m_e}{en}\left(\frac{\partial\vec{J}}{e\partial t}+\nabla\cdot n\left(\vec{v}_i\vec{v}_i-\vec{v}_e\vec{v}_e\right)\right),
\end{equation}

\noindent where $e$ is the elementary charge, $m_e$ is the electron mass, $n$ is the plasma number density, $\vec{v}_i$ is the ion bulk velocity, and $\bar{P}_e$ is the electron pressure tensor \cite{Torbert.2016b}. Both studies found that the reconnection electric field at the X-point was balanced by the electron inertia term $-\frac{m_e}{en}\nabla\cdot n(\vec{v}_e\vec{v}_e)$ and that the reconnection electric field at the electron stagnation point was governed by the divergence of the off-diagonal (agyrotropic) elements of the electron pressure tensor. They questioned whether reconnection was then a fundamentally reversible process, given that bulk inertial effects appeared to dominate at the reconnection X-point.

\cite{Torbert.2016b} was the first to calculate the electron inertia and pressure divergence terms of (1) with MMS data. They determined that, for the anti-parallel asymmetric EDR of \cite{Burch.2016b}, the energy conversion near the electron stagnation point was driven by both pressure divergence and electron inertia at a ratio of $\sim$4:1. They also found that the error in the gradient approximation (and/or the anomalous resistivity) was considerable. \citeauthor{Torbert.2016b} supported their findings by analyzing a 2.5-d PIC simulation of the event, which found these two terms were driving energy conversion at about the same ratio and that the anomalous resistivity term was negligible. \textit{Rager et al.,} [submitted] analyzed the same event as \cite{Torbert.2016b} and found that the gradients in the perpendicular ($\sim M-L$) elements on the pressure tensor were dominant. However, they concluded that the terms in Ohm's law could not be fully accurately resolved, even with their higher time resolution (7.5 ms) electron data.

\subsection{Manuscript organization}

In this study, we (1) introduce an EDR event that was observed by all four MMS spacecraft during an orbit where the tetrahedron had a very small (6.4 km) inter-probe separation, (2) analyze the electron dynamics within the diffusion region, and (3) analyze the form of generalized Ohm's law near the reconnection X-point, with specific focus on the role of electron agyrotropy and 3-d (out-of-plane) structure. We find that the small inter-probe separation allows us to estimate the terms in Ohm's law with greater accuracy than what has previously been reported. Furthermore, we find that the local solution to Ohm's law differs significantly from the predictions of 2.5-d PIC simulations, where $\vec{J}\cdot\vec{E}'$ may have been enhanced by turbulent, 3-d structure of the current layer. These findings are relevant to open questions about the nature of energy conversion in asymmetric EDRs, specifically regarding the influence of local fluctuations in the current sheet structure on dissipation.

The following section describes the data used in this investigation and the coordinates used to organize the data. Section 3 provides an overview of the magnetopause crossing during which the EDR was detected and an analysis of the long-duration magnetosphere-side separatrix. Section 4 analyzes the electron velocity distributions near the X-point. Section 5 analyzes the terms in generalized Ohm's law during the X-point crossing. Finally, the results are summarized, discussed, and compared/contrasted with results from a similar event in Section 6.

\section{Data, analysis methods, and LMN coordinates}

We use the highest possible resolution data from the four MMS spacecraft on 28 November 2016. Moments and VDFs for electrons are obtained by the fast plasma investigation (FPI) once per 30 ms \cite{Pollock.2016}. Ion moments and VDFs are obtained by FPI once per 150 ms. Measurements of the DC magnetic field are obtained by the fluxgate magnetometers (FGM) at 128 vectors per second \cite{Russell.2016}. Measurements of the AC magnetic field are obtained by the search coil magnetometers (SCM) at 8196 vectors per second \cite{LeContel.2016}. The coupled AC and DC electric field vector is measured by the electric field double probes (EDP) at 8196 vectors per second \cite{Ergun.2016,Lindqvist.2016}. We use the level 2 (l2) data from the FPI, FGM, and SCM, all of which are publicly available at the MMS science data center. We use the better calibrated l3 electric field data, which are available upon request. 

To calculate the pressure divergence and inertial terms in Ohm's law, we assume that, within the volume of the spacecraft tetrahedron, gradients in the fields and plasma moments are approximately linear \cite{ISSIchap14}. A sliding overlapping boxcar scheme is used to smooth the coupled AC/DC electric field vector. When the electric field is used to calculate $\vec{J}\cdot\vec{E}'$, we use a boxcar width of 30 ms, chosen to match the cadence of FPI electron measurements. 

A magnetopause-normal LMN coordinate system was determined by applying the minimization of Faraday residue (MFR) technique \cite{KhrabrovandSonnerup.1998} to burst-mode magnetic and electric field data from MMS1 during a 28-second interval starting at 07:36:32 UT. Here, $L$ is the direction of maximum magnetic shear, $N$ is the magnetopause normal, and $M$ completes the right-handed coordinate system. In the geocentric solar ecliptic (GSE) coordinate basis, $L$, $M$, and $N$ are [0.177617, --0.158717, 0.971216], [0.244648, --0.948804, --0.199795], and [0.953205, 0.273093, --0.129694], respectively. The average velocity of the magnetopause was --11.4 km/s $\hat{N}$, as determined with MFR. Due to the long duration of the magnetopause crossing ($\sim$30 seconds), it is possible that the configuration of the magnetopause changed over the course of the event. It is also possible that kinetic-scale instabilities cause the local current and boundary orientations to differ from the larger-scale configuration \cite{Ergun.2016b,Ergun.2017,Price.2016,Price.2017}. Four-point timing analysis \cite{ISSIchap10} of the $B_{ZGSE}$ reversal point (at approximately 07:36:55 UT) yields a normal vector of [0.98011, 0.192104, --0.049797], again in GSE, and a boundary velocity of --31 km/s $\hat{N}$. These two normal vectors, one determined with MFR and the other from timing analysis, differ by approximately 10$^\circ$. There is a significant difference in the two boundary velocities, as is discussed in the next section. A third boundary-normal system, local to the X-point and referred to as LMN-X, is discussed in Section 4. The normal direction in LMN-X differs by less than 3$^\circ$ with the normal determined by MFR and by less than 7$^\circ$ with the normal direction from timing analysis. 

\section{Large-scale observations during the magnetopause crossing} 

\begin{figure*}
\noindent\includegraphics[width=39pc]{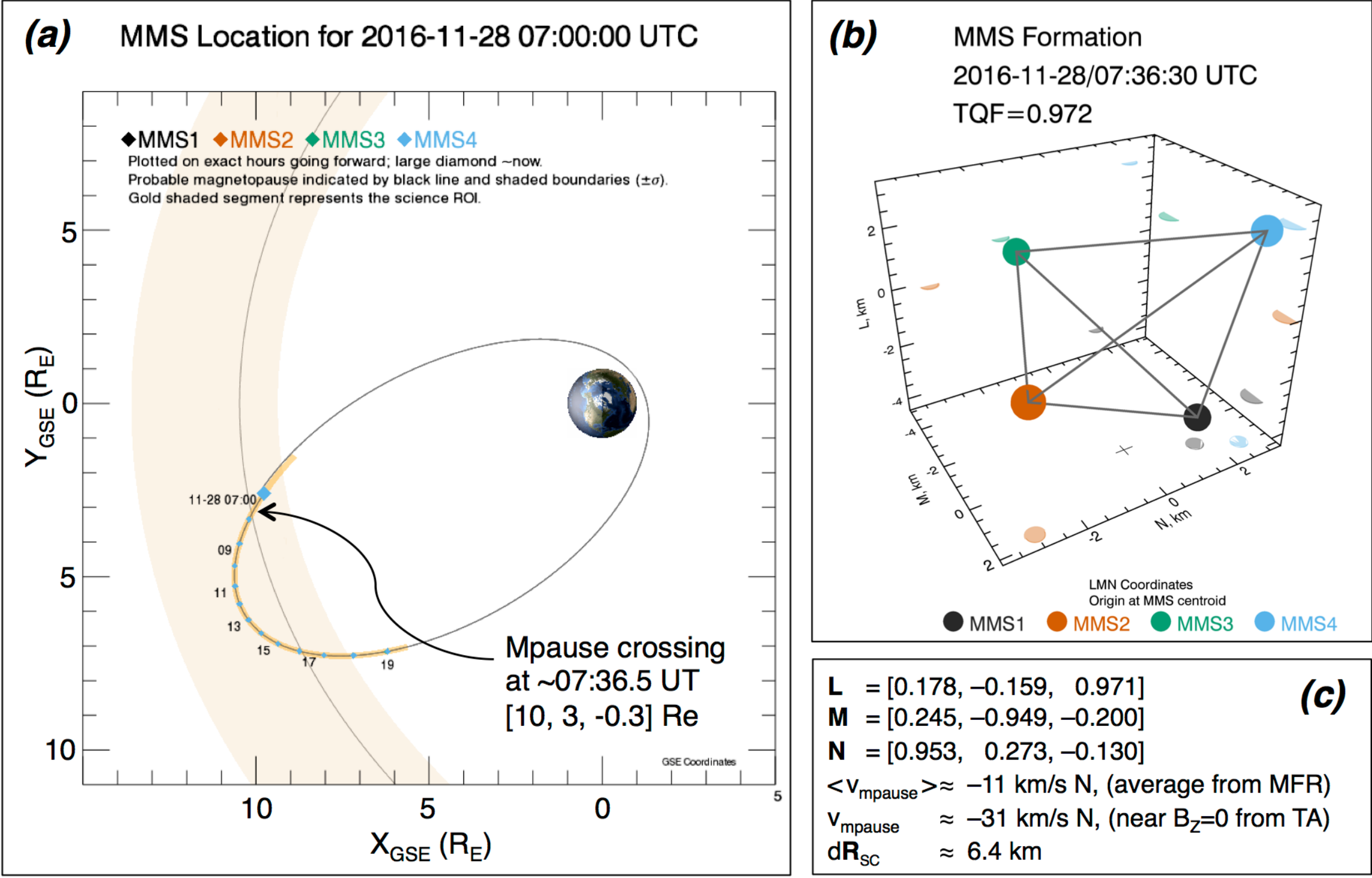}
\caption{(a) Position of the MMS constellation relative to the nominal position of the magnetopause, as predicted by the empirical model of \cite{Shue.1998}. (b) Formation of the tetrahedron in LMN coordinates (see Section 2). (c) Relevant data, including the LMN coordinate axes in GSE, the average velocity of the magnetopause determined with MFR of data from the full $\sim$30-second crossing, the comparatively instantaneous velocity of the magnetopause determined by timing analysis of the magnetic field vector near the $B_Z$ reversal point, and the electron inertial length $d_e$ in the magnetosheath. Panel (a) is taken from the quicklook orbit plot archive on the MMS science data center.}
\label{mec}
\end{figure*}

On 2016-11-28 at approximately 7:36:30--7:37 UT, MMS crossed the magnetopause duskward of the subsolar point near the GSE equatorial plane (Figure \ref{mec}a). The spacecraft separation was very small, with an average inter-probe separation of 6.41 km $\pm$0.50 km (Figure \ref{mec}b). The crossing was directed from the magnetosphere outward into the magnetosheath and MMS obtained measurements in both inflow regions, i.e., the magnetosphere proper and magnetosheath proper. The asymptotic upstream conditions are listed in Table 1, where the magnetosphere-side and magnetosheath-side parameters were determined between 7:35:10--7:36:10 UT (magnetosphere proper) and 7:37:05--7:38:05 UT (magnetosheath proper), respectively. Hereafter, the subscripts ``\textit{sh}'' and ``\textit{sp}'' are used to describe parameters from the magnetosheath and magnetosphere, respectively.

\begin{table}
\centering
\caption{Asymptotic upstream conditions and parameters related to asymmetric reconnection determined from MMS1 data.}
\begin{tabular}{| l | r | r | c | l | r |}
\cline{1-3}\cline{5-6}
& \multicolumn{1}{|c|}{Sphere} & \multicolumn{1}{|c|}{Sheath} & \hspace{3mm} &  \multicolumn{2}{|c|}{Boundary params.} \\
\cline{1-3}\cline{5-6}
\cline{1-3}\cline{5-6}
\multirow{3}{*}{$\left< \vec{B} \right>$ [nT]} & 47.65 $\hat{L}$ & --20.61 $\hat{L}$  & \cellcolor[gray]{0.8} & $|B_{L,sh}/B_{L,sp}|$ & 0.43 \\ 
\cline{2-3}\cline{5-6}
                                           & --8.08 $\hat{M}$ &  --18.16 $\hat{M}$ & \cellcolor[gray]{0.8}& $|\left<B_M\right>/B_{L,sh}|$ & 0.64 \\
\cline{2-3}\cline{5-6}
                                           &   8.08 $\hat{N}$ & --0.58 $\hat{N}$      & \cellcolor[gray]{0.8}& Shear angle & 128.6$^\circ$ \\
\cline{1-3}\cline{5-6}
$\left<|\vec{B}|\right>$ [nT] & $49.00\pm0.67$ & $27.46\pm2.81$    & \cellcolor[gray]{0.8}& $n_{sh}/n_{sp}$ & 26.5 \\
\cline{1-3}\cline{5-6}
$\left<\mathrm{cos}^{-1}\left(\left<\hat{b}\right>\cdot\hat{b}\right)\right>$ & $1.44^\circ$ & $7.18^\circ$ & \cellcolor[gray]{0.8}& $v_{A,asym}\tablenotemark{a}$  [km/s] & 210 \\
\cline{1-3}\cline{5-6}
$\left<n\right>$ [cm$^{-3}$] & $0.57\pm0.17$ & $14.99\pm1.31$  & \cellcolor[gray]{0.8}& $T_{tot, asym}\tablenotemark{b}$  [eV] & 431 \\
\cline{1-3}\cline{5-6}
$\left<T_e\right>$ [eV]       & $75.0\pm19.6$  & $42.5\pm2.67$    & \cellcolor[gray]{0.8}& $v_{s,asym}\tablenotemark{c}$  [km/s] & 203 \\
\cline{1-3}\cline{5-6}
$\left<T_i\right>$  [eV]       & $3711\pm1721$ & $333.3\pm23.2$  & \cellcolor[gray]{0.8}& $d_{i,asym}$ ($d_{e,asym}$)$\tablenotemark{d}$ [km] & 69.9 (1.63) \\
\cline{1-3}\cline{5-6}
\end{tabular}
\tablenotetext{a}{Asymmetric Alfven speed \cite{CassakandShay.2007}.}
\tablenotetext{c}{Asymmetric sound speed, based on the asymmetric total temperature $^\mathrm{b}$ $T_{tot,asym}$ \cite{Shay.2014,Cassak.2017}.}
\tablenotetext{d}{Asymmetric ion (electron) inertial lengths \cite{CassakandShay.2009}.}
\end{table}

\begin{figure*}
\noindent\includegraphics[width=39pc]{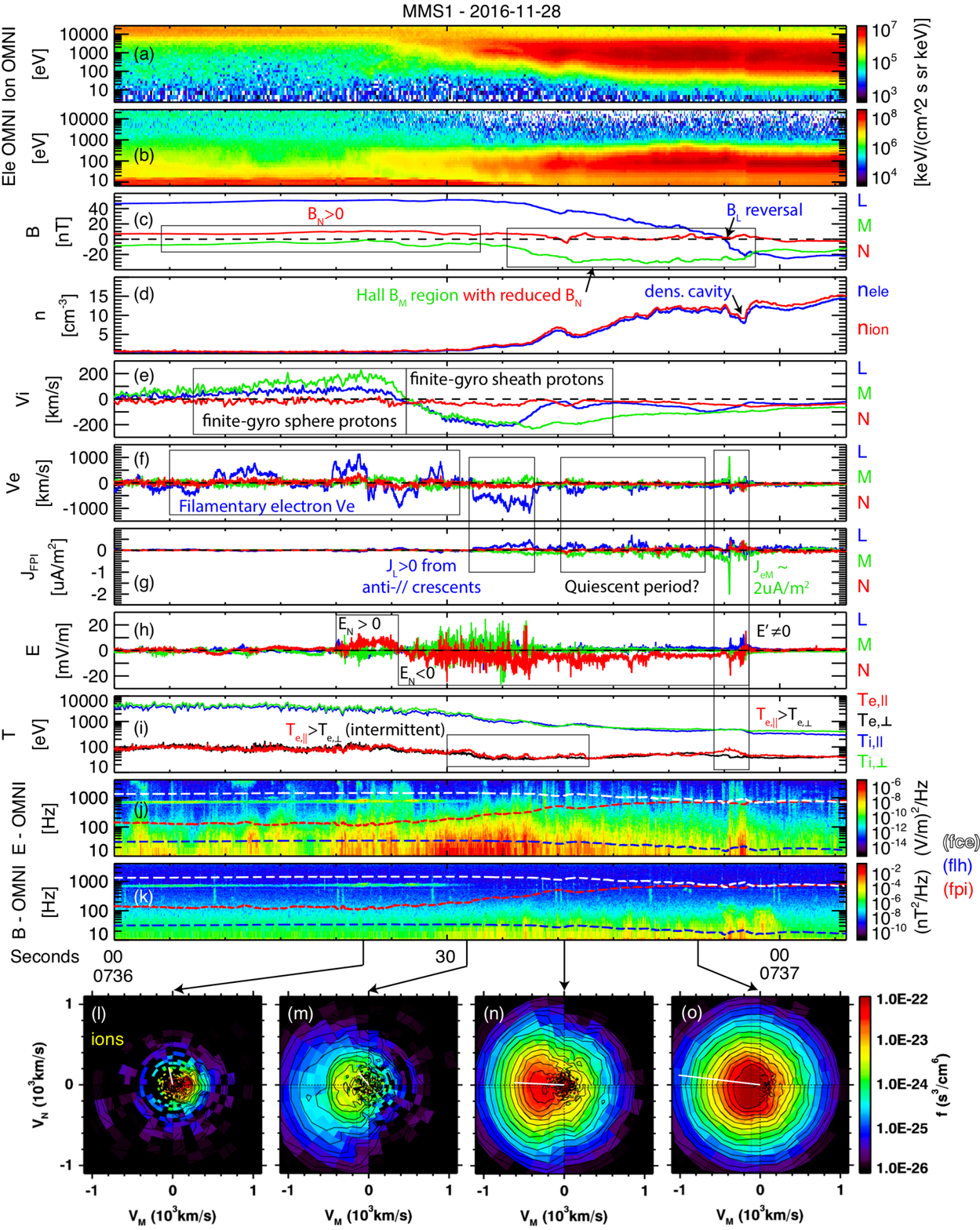}
\caption{Omnidirectional (a) ion and (b) electron spectrograms, (c) magnetic field vector, (d) ion and electron number densities, (e) ion and (f) electron bulk velocities, (g) current density vector calculated from the plasma moments, (h) electric field vector in the spacecraft frame, (i) ion and electron temperatures, and omnidirectional (j) electric and (k) magnetic field power spectrograms. The white, blue, and red traces on panels (j) and (k) are the electron-cyclotron, lower-hybrid, and plasma-ion frequencies, respectively. (l-o) 0.9-second averages of the ion VDF.}
\label{MPause_full}
\end{figure*}

Time-averaged MMS1 data from the magnetopause crossing is shown in Figure \ref{MPause_full}. The shear angle was approximately 129$^\circ$. There was a strong asymmetry in $B_L$ ($B_{L,sp}/B_{L,sh}\approx0.4$), an average guide field approximately half the size of the magnetosheath reconnecting field ($\left<B_M\right>/B_{L,sh}\approx0.6$), and strong asymmetries in the temperatures ($T_{e,sh}/T_{e,sp}\approx0.6$, $T_{i,sh}/T_{i,sp}\approx0.09$) and density ($n_{sh}/n_{sp}\approx27$). The positive normal $B_N$, which was observed throughout the crossing, and the strong negative Hall $B_M$, which was observed in the magnetopause plasma mixing region, both indicate that MMS crossed the magnetopause southward of an X-line. There was also a considerable asymmetry in the ion thermal pressure of $P_{i,sh}/P_{i,sp}\approx$2.3, which was predicted to be the source of free energy for the drift instability by \cite{Price.2017}. 

\begin{figure*}
\noindent\includegraphics[width=39pc]{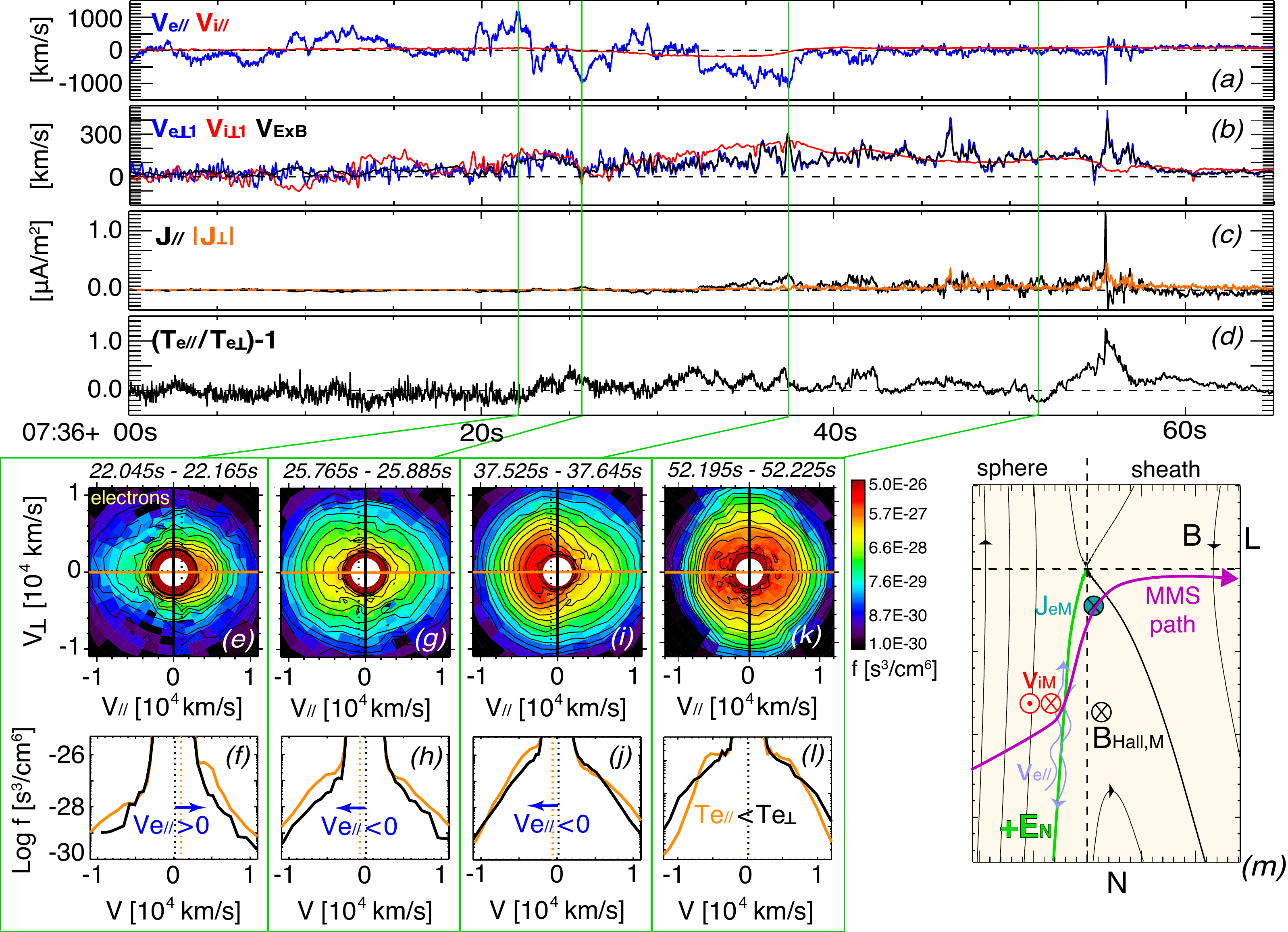}
\caption{Plasma data from MMS1 in local (time-dependent) B-field coordinates. (a) Parallel and (b) perpendicular velocities of electrons (blue) and ions (red). Also in (b), the $\vec{E}\times\vec{B}$-drift velocity (black). (c): The parallel (black) and perpendicular (orange) current density. (d) The electron anisotropy. Vertical lines mark the times where (e)-(l), 2-d and 1-d cuts of eVDFs, were measured. (m) A schematic diagram of the estimated path of MMS through the reconnection region with some of the observed reconnection signatures.}
\label{vbplot}
\end{figure*}

The transition from a magnetosphere-like ion population (low density with a $\sim$several keV component and a cold $\sim$hundreds eV component) to a magnetosheath-like ion population (higher density, $\sim$several hundred eV) began at approximately 7:36:27 UT and coincided with reversals in the $L$ and $M$ components of the ion bulk velocity and a reversal in the normal electric field. In Figure \ref{MPause_full}h, the reversal of $E_N$ occurs near 7:36:26 UT and is consistent with a crossing of the magnetosphere-side separatrix given the predicted path of MMS shown in Figure \ref{vbplot}m \cite{Malakit.2013,Shay.2016}. In Figure \ref{MPause_full}e, this reversal of $v_{iM}$ occurs at the boundary between the boxes labeled ``finite-gyro sphere protons'' and ``finite-gyro sheath protons''. The second of the two $\vec{v}_{i}$ peaks is caused by the higher density magnetosheath protons penetrating across the magnetopause and completing half of a gyro-orbit \cite{Phan.2016them}, which is consistent with the observed dispersion (higher-energy larger-gyro-radius sheath protons penetrate deeper into the magnetosphere, appearing in Figure \ref{MPause_full}a earlier than the lower-energy sheath protons). This explanation for the $v_{iM}$ reversal is also consistent with the evolution of the ion distribution function between Figures \ref{MPause_full}o and m, where the ions are largely gyrotropic near the field reversal region (panel o) and become agyrotropic further into the magnetosphere, until eventually only sheath ions with perpendicular velocities tangential to the magnetopause are observed (panel m). The first of the two $v_{iM}$ peaks, which had $v_{iM}>0$, is a result of high-energy magnetospheric ions with guiding centers nearer the magnetopause being lost into the magnetosheath. A simple schematic of this effect for a 180$^\circ$-shear (fully anti-parallel) boundary is found in Figure 2b of \cite{Phan.2016them}. In the case of this 2016-11-28 event, where the shear angle is less than $180^\circ$, the large $v_{iL}$ that accompanies the large $v_{iM}$ may result from the rotation of the Lorentz force direction in $M$ by the moderate guide field. The strong $v_{iL}$ is not likely associated with an ion jet, as the rotation of $v_{iL}$ is observed before the separatrix, as discussed later in this section.

The reversal of $v_{iM}$ is observed roughly 320 km (4.6 $d_{i,asym}$) magnetosphere-ward of the $B_L$ reversal, given a timing difference of 28 seconds and an average magnetopause normal velocity of --11.4 km/s, which was determined by MFR analysis (see Section 2). This estimated distance is comparable to the effective ion thermal gyro-radius of 305 km, based on the magnetosheath ion thermal velocity and the asymmetric ion gyro-frequency. Simulations of 2-d anti-parallel reconnection predict that this finite-gyro-radius effect occurs within a region roughly $\pm15$ $d_i$ downstream of the X-line \cite{Shay.2016}, and observations of this finite-gyro-radius effect have been used as evidence of close proximity to the X-point \cite{Khotyaintsev.2016,Phan.2016them}. 

During the rotation of $v_{iM}$, roughly between 7:36:05--30s UT, field-aligned streaming (upward, towards the X-point) and anti-field-aligned counter-streaming (downward, away from the X-point) electrons were observed with speeds up to $\sim$500-1000 km/s (Figure \ref{vbplot}a and Figure \ref{MPause_full}f). Filamentary electron velocity enhancements have been previously identified in the MMS data \cite{Phan.2016,WangRS.2017}. Figures \ref{vbplot}e-f and \ref{vbplot}g-h show two cuts of the eVDF taken during intervals of streaming and counter-streaming low-density electrons. The upward-moving electrons, in general, were observed at lower energies ($\sim20$ eV) and lower densities ($\sim$0.3 cm$^{-3}$), whereas the downward-moving electrons were observed at higher energies ($\sim100eV$), with slightly higher densities ($\sim$0.5 cm$^{-3}$) and a larger anisotropy (seen in Figure \ref{vbplot}d and in the comparison of Figures \ref{vbplot}f and h). In the 2.5-d PIC simulation of \cite{Shay.2016}, weak upward and strong downward parallel electron velocities were found to straddle the southern magnetosphere-side separatrix, where the magnetic field lines transition from closed to opened. In this picture, illustrated in Figure \ref{vbplot}k, the downward-moving electrons lie on reconnecting or recently reconnected field lines. Further evidence for the $\vec{v}_{e\parallel}$ reversal lying on or near the open-closed boundary is shown in Figure \ref{vbplot}i, which shows that the last and strongest interval of counter-streaming electrons (near 7:36:35 UT) contained outflow crescent electrons \cite{Khotyaintsev.2016,Hwang.2017,Shay.2016}. Anisotropic but gyrotropic ``outflow'' crescents, of the type shown in Figure \ref{vbplot}i, are expected up to several $d_i$ downstream of the center of the EDR \cite{Shay.2016}.

The multiple reversals of $\vec{v}_{e\parallel}$ seen in Figure \ref{vbplot}a may indicate that MMS crossed the magnetosphere-side separatrix, returned back, then crossed again, perhaps multiple times. No similar signature of this multiple-crossing-type motion is seen in the single clean rotation of $\vec{v}_i$, which may indicate that the electron separatrix layer was moving in and out relative to the ion-scale boundary. The magnetic field data shown in Figure \ref{MPause_full}c may support this theory, as $|B_M|$ ($B_L$) is slightly enhanced (reduced) during the times with downward-moving electrons, as would be expected for a crossing into the separatrix. Electromagnetic whistler waves with frequencies of half the electron cyclotron frequency, as are seen in Figures \ref{MPause_full}j and k, are predicted to form in the inflow region near the separatrix \cite{Fujimoto.2014}. The persistence of the whistlers may indicate that MMS remained very near the separatrix for some time before finally crossing the boundary at roughly 7:36:32 UT. Waves near/below the lower-hybrid frequency are also observed throughout the separatrix crossing and also near the $B_L$ reversal. Electrostatic waves of similar frequencies and locations have been reported in a number of EDR events, and were associated with drift-instability-driven corrugations of the magnetopause current layer \cite{Ergun.2016b,Ergun.2017,Price.2016,Price.2017}. The final crossing of the separatrix occurred roughly 10 seconds after the reversal of $v_{iM}$, which implies that the thickness of the reconnection boundary layer is below the magnetosheath proton gyro-scale (given that the reversal of $v_{iM}$ is caused by penetrating sheath protons). 

After the final crossing of the magnetosphere-side separatrix near 7:36:35 UT, the reconnecting magnetic field component $B_L$ began a $\sim$20-second and nearly monotonic decrease, excluding small fluctuations. The strength of the out-of-plane magnetic field $B_M$ also increased substantially, as did the plasma density, whereas the normal magnetic field component $B_N$ decreased. Between $\sim$7:36:40--50s UT, labeled as a ``Quiescent period'' on Figure \ref{MPause_full}, there are very few particle signatures that indicate proximity to an X-point (e.g., super-Alfvenic electron jets, electron crescent distributions, etc.), other than the presence of demagnetized protons (Figure \ref{vbplot}b). We have labeled this period as ``quiescent'' in reference to the description by \cite{BurchandPhan.2016} of their moderate guide field EDR event, as they also found a period with little electric field activity and a reduced temperature anisotropy between the field reversal point and the normal flow reversal point. In our quiescent period, there are no obvious $L$-directed electron jets, as have been observed between the X-point and separatrix for anti-parallel reconnection events \cite{Hwang.2017}. The electrons also appear magnetized during this period, with an average perpendicular bulk flow ($\sim150$ km/s) below the asymmetric Alfven speed (210 km/s, Table 1). 

The quiescent period ends where the focus of the remainder of this investigation begins, at approximately 7:36:50 UT near the reversal of $B_L$. Many signatures of the inner diffusion region are observed in this region, including an intense and narrow out-of-plane electron current, an intense broad region of electron anisotropy with $T_{e,\parallel}<T_{e,\bot}$ followed by $T_{e,\parallel}>>T_{e,\bot}$ (Figure \ref{vbplot}m-l), a moderate narrow region of electron agyrotropy, $\vec{J}\cdot\vec{E}'>0$, etc., all of which are discussed in the next section. {As is shown in Figure \ref{vbplot}c, the current near the $B_L$ reversal is largely field aligned, which is common in MMS observations of asymmetric guide-field reconnection \cite{BurchandPhan.2016,Eriksson.2016,Genestreti.2017,Ergun.2017,Chen.2017}.} It is unlikely that MMS crossed the center of the reconnection site, or the exact X-point, since the Hall field $B_M<\left<B_{M,guide}\right>$ remained large throughout the $B_L$ reversal interval. The magnetosheath-side separatrix was observed approximately 1.4 seconds after the large out-of-plane current, which, using the boundary velocity of -31 km/s $\hat{N}$ determined from timing analysis of $B_L=0$, corresponds to a distance of 43 km ($\sim0.6$ $d_{i,asym}$, where the asymmetric or hybrid inertial length $d_{asym}$ is defined in \cite{CassakandShay.2009}). For comparison, the separation between the $B_L$ reversal and the magnetosphere-side separatrix was estimated to be roughly 300 km ($\sim4.3$ $d_{i,asym}$). This asymmetry in the observed half-thickness of the reconnection layer (thicker on the magnetosphere side and thinner on the magnetosheath side) may support the path of MMS drawn in Figure \ref{vbplot}m, where MMS entered the thicker outer diffusion region and exited the thinner inner diffusion region. After roughly 7:37 UT, following the sheath-side separatrix crossing, MMS entered and remained in the magnetosheath proper for several minutes.

In summary, we have found the following list of signatures, many of which are associated with kinetic-scale reconnection:
\begin{enumerate}
\item{Finite-gyro-radius penetration of magnetosheath protons into the magnetosphere-side inflow region, beyond the electron-scale separatrix (see reversal in $v_{iM}$ at approximately 7:36:27 UT in Figure \ref{MPause_full}),}
\item{Intense parallel electron flows and anti-parallel flows of ``outflow'' crescent electrons in the magnetosphere-side separatrix,}
\item{Intermittent electron anisotropy with $T_{e,\parallel}/T_{e,\bot}\sim$1.5 near the magnetosphere-side separatrix,}
\item{Electromagnetic whistler waves near the magnetosphere-side separatrix and electric field waves below $f_{L-H}$ near the separatrix and $B_L$ reversal,}
\item{A ``quiescent'' region with unmagnetized ions and $\vec{E}\times\vec{B}$-drifting electrons between the magnetosphere-side separatrix and the $B_L$ reversal,}
\item{A reduction in $B_N$ and drastically different distances between the two separatrices and the $B_L$ reversal (see diagram in Figure \ref{vbplot}m), which are consistent with an approach toward the X-point,}
\item{Signatures of the inner diffusion region (next section) near the $B_L$ reversal and intense out-of-plane current ($\sim$7:36:55), and}
\item{Signatures of the magnetosheath-side separatrix (7:36:57) following the observation of the electron-scale $J_M$ near the X-point and preceding entry into the magnetosheath proper.}
\end{enumerate}

Based on these findings, we conclude that MMS crossed the magnetosphere-side separatrix perhaps on the order of $<$10--15$d_i$ away from the center of the EDR. Then, MMS remained in the reconnection region while traveling northward towards the X-point, before exiting the magnetopause near the central EDR. This path is illustrated in Figure \ref{vbplot}m.

\section{Electron dynamics near $B_L$ reversal}

\begin{figure*}
\noindent\includegraphics[width=35pc]{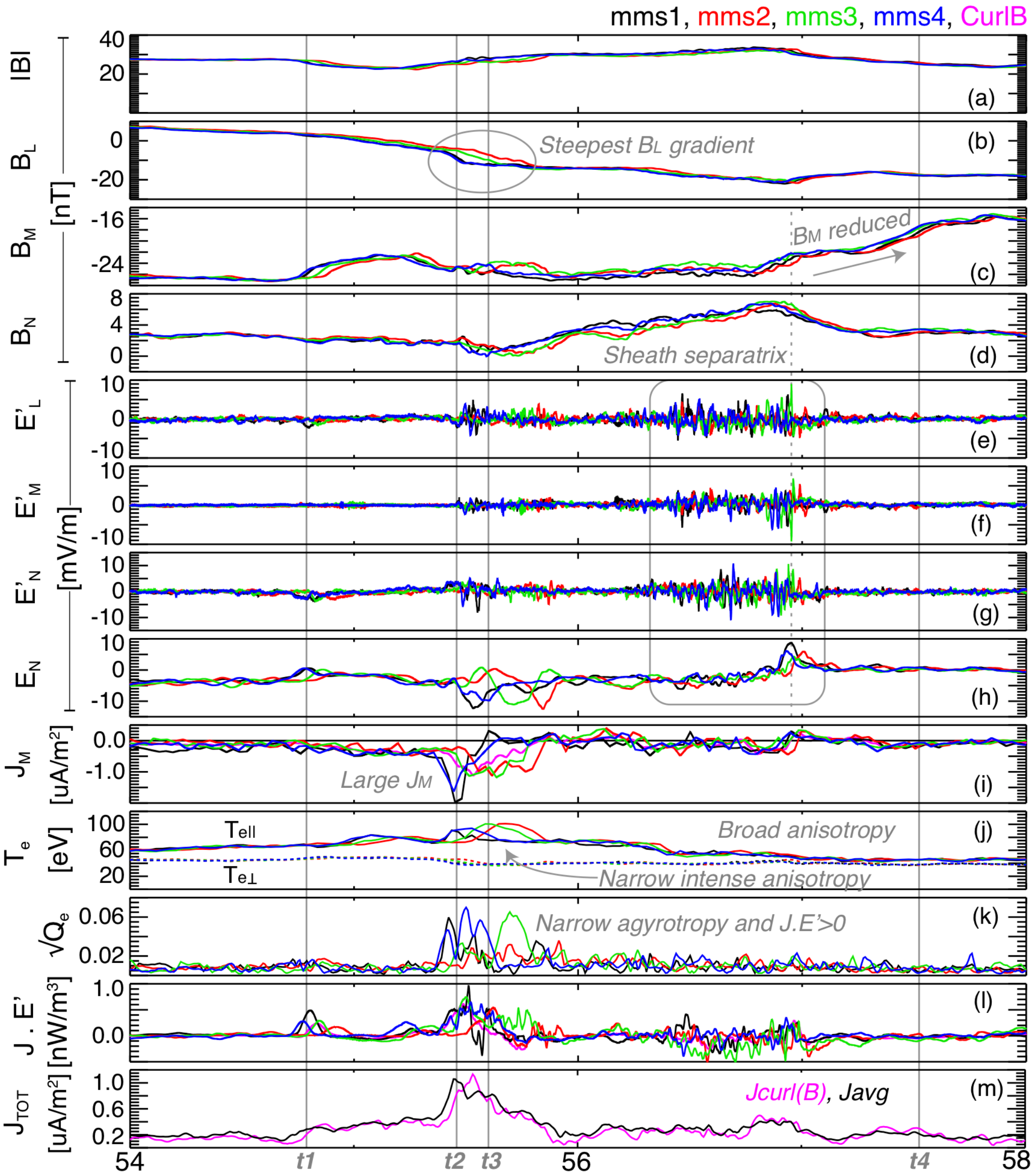}
\caption{Data from the 4 spacecraft near the X-point, including (a) the total magnetic field strength, (b)--(d) the $L$, $M$, and $N$ components of the magnetic field, (e)--(g) the three components of the electric field in the electron frame, (h) the normal electric field in the spacecraft frame, (i) the out-of-plane current density, (j) the electron temperatures, (k) the electron agyrotropy, (l) the electron-frame energy conversion rates, and (m) the total current density from (magenta) the curlometer method and (black) the average of the FPI plasma moments.}
\label{4probe_plot}
\end{figure*}

\begin{figure*}
\noindent\includegraphics[width=39pc]{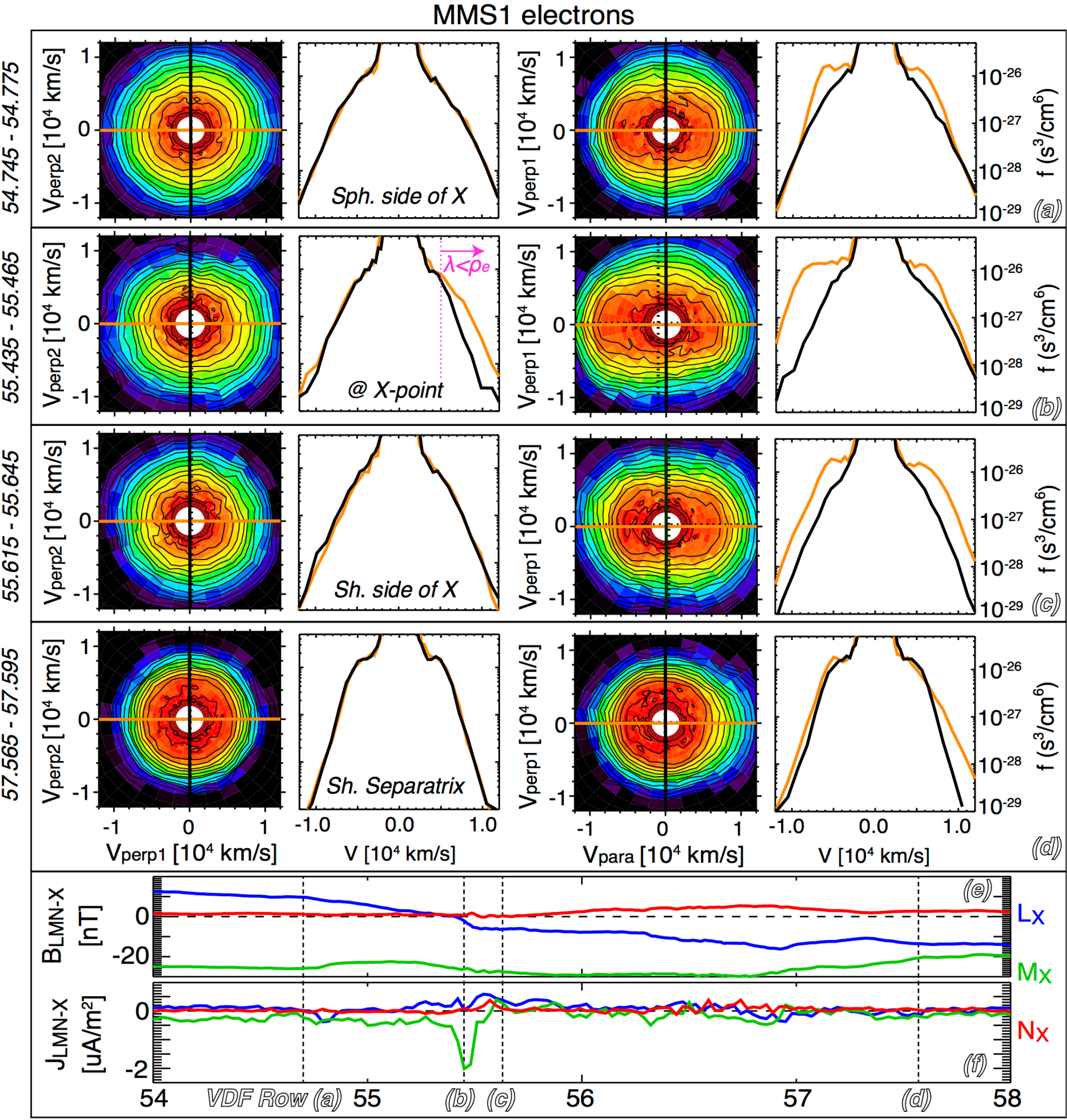}
\caption{Data from MMS-1 near the X-point. (a)--(d): first column shows 2-d cuts of the eVDF in the $v_{\bot1}-v_{\bot2}$ plane, second column shows 1-d cuts along $v_{\bot1}$ (orange) and $v_{\bot2}$ (black), third column shows 2-d cuts in the $v_{\parallel}-v_{\bot1}$ plane, and final column shows 1-d cuts along $v_{\parallel}$ (orange) and $v_{\bot1}$ (black). The color bars are identical to those in Figure \ref{vbplot}. The times where the four eVDFs (a--d) were measured are marked on panels (e--f) with vertical lines. (e) Magnetic field and (f) current density in the LMN-X system, where $L_X$=[0.157, 0.035, 0.987], $M_X$=[0.240, $-$0.971, $-$0.0039], and $N_X=L_X\times M_X$.}
\label{Xmms1vdf}
\end{figure*}

Figure \ref{4probe_plot} shows data from the four spacecraft around the $B_L$ reversal, where the largest out-of-plane current (Fig \ref{4probe_plot}i) is observed at 7:36:55.5 UT. The vertical grey lines in Figures \ref{4probe_plot} and \ref{Xmms1vdf}e-f mark the times of the eVDFs shown in Figure \ref{Xmms1vdf}a-d. MMS1 and 4, which crossed the current layer nearly simultaneously, both observed $J_M\approx-2$ $\mu$A/m$^2$ for approximately 0.15 seconds, which, given a boundary speed of 31 km/s, corresponds to a thickness of approximately 5 km, or 3 $d_{e,asym}$, or 0.07 $d_{i,asym}$. MMS3, which crossed after 1 and 4, and MMS2, which crossed last, both observed a $J_M$ layer approximately half as intense and twice as thick as the current layer observed by MMS1 and 4. The average current sheet thickness was determined from the curlometer current (shown in magenta in Figure \ref{4probe_plot}i) to be roughly 9 km. This thickness is significantly larger than the local electron inertial length or gyro-radius ($\rho_e\sim$0.6 km, $d_e\sim$1.3 km) and is closer to the scale suggested by \cite{Price.2017} for turbulent reconnection $\sqrt{\rho_e\rho_i}\approx13.5$ km. 

The current sheet thickness is also close to the average inter-spacecraft separation, 6.4 km, which means that the linear gradient assumption should be applied with caution. However, since (a) the curlometer current very closely resembles the average of the currents measured by each of the four spacecraft (Figure \ref{4probe_plot}m) and (b) the linear-approximate ratio of $\nabla\cdot\vec{B}/\nabla\times\vec{B}$ is small, approximately 10\% at the center of the current sheet, we conclude that the linear approximation does a reasonable job of estimating the current from the curl of $\vec{B}$. 

A broad temperature anisotropy of $T_{e,\parallel}/T_{e,\bot}\approx2$ surrounds the electron $J_M$ layer. An extreme and narrow anisotropy of $T_{e,\parallel}/T_{e,\bot}\approx3.5$ is  colocated with the $J_M$ layer. There is also a moderate electron agyrotropy observed along with the large $J_M$, which, is described by the $\sqrt{Q}$ parameter of \cite{Swisdak.2016} in Figure \ref{4probe_plot}k. A $\sqrt{Q}$ of 0 corresponds to a fully gyrotropic distribution, where $P_{e,\bot1}=P_{e,\bot2}$ and all off diagonal elements of the pressure tensor are zero. A $\sqrt{Q}$ of 1 corresponds to a fully agyrotropic distribution. For reference, the maximum value of $\sqrt{Q}$ for the 2015-12-08 guide-field central magnetopause EDR event of \cite{BurchandPhan.2016} was $\sqrt{Q}=0.09$. 

During the period with the largest agyrotropy, all four MMS spacecraft also detected mostly positive $\vec{J}\cdot\vec{E}'$. The rate of energy conversion is much weaker than others have reported in the central EDR. For example, the moderate guide field event of \cite{BurchandPhan.2016} and the nearly anti-parallel event of \cite{Burch.2016b} had, respectively, electron-frame energy conversion rates 10 and 20 times larger than what is observed here. For anti-parallel asymmetric reconnection, the largest $\vec{J}\cdot\vec{E}'>0$ is expected near the electron stagnation point where agyrotropic electron crescent VDFs are observed. However, recent work showed that the $\vec{J}\cdot\vec{E}'>0$ region is shifted towards the X-point for guide-field asymmetric reconnection \cite{Genestreti.2017,Cassak.2017}.

Figure \ref{Xmms1vdf} shows four eVDFs from MMS1 taken near the X-point. The four rows correspond to the four times marked by solid vertical lines in Figure \ref{4probe_plot}. The first two columns show that the electrons remain largely gyrotropic throughout the interval, except at the narrow region with $J_M<<0$ and $\vec{J}\cdot\vec{E}'>0$ (\ref{Xmms1vdf}b). The agyrotropy is visible as an expansion (rather than a displacement) of the distribution function in the $\vec{v}_{\bot1}>0$ direction. 

As is shown in Figure \ref{4probe_plot}, the peaks of $\sqrt{Q}$, $\vec{J}\cdot\vec{E}'>0$, and $J_M$ are observed approximately 0.5 seconds after the $B_L$ reversal. From \cite{Hesse.2014,Hesse.2016}, we expect that the largest electron agyrotropy should occur where the local electron gyro-radius $\rho_{e}$ approaches or exceeds the local magnetic scale size $\lambda=B_L/\mu_0J_M$, given that this is where electrons observe a considerably different $B_L$ on either side of their orbits. Based on the data in Figure \ref{4probe_plot}, this condition is only met near the $B_L$ reversal, where the electrons are largely gyrotropic. We conclude that the LMN system determined for the entire magnetopause crossing does not organize the data near the X-point. Local LMN coordinates are therefore determined by applying minimum variance analysis to the current density vector measured by MMS1 between 52--59 s. In this ``LMN-X'' coordinate system, in which data in Figure \ref{Xmms1vdf}e-f are showed, the $B_{LX}$ reversal occurs with the peak of $J_{MX}$ and the condition $\lambda\leq \rho_{e}$ is met for the period of maximum agyrotropy for electrons with energies exceeding $\sim$80 eV. (Note that the ``LMN-X'' coordinates do not organize the larger-scale magnetopause crossing data as well as our previously-defined LMN system.)

Cuts of the eVDF within the narrow electron current layer are shown in Figure \ref{Xmms1vdf}b. The distribution in \ref{Xmms1vdf}b resembles adjacent eVDFs, where the phase space density in the $\vec{v}_{\parallel}<0$ direction is significantly greater than in the $\vec{v}_{\parallel}>0$ direction, as is consistent with the direction of the strong out-of-plane parallel current. Electrons with this type of $\vec{v}_{\parallel}-\vec{v}_{\perp}$ distribution have been observed in a very low-shear central EDR \cite{Eriksson.2016} and in a number of moderate-shear central EDRs \cite{Genestreti.2017}. \cite{Genestreti.2017} suggested that this type of distribution is typical of guide-field X-points, and that the acceleration and heating of the inflowing magnetosheath electrons along the guide field is tied to the large  colocated $\vec{J}\cdot\vec{E}'>0$. The agyrotropic $(\vec{v}_{\bot1}-\vec{v}_{\bot2})$ structure of the eVDF in the current layer is most pronounced for higher velocities, above roughly 5,000 km/s (80 eV), which is the same energy range for which $\lambda\leq \rho_{e}$.

\section{Energy conversion near $B_L$ reversal}

\begin{figure*}
\noindent\includegraphics[width=39pc]{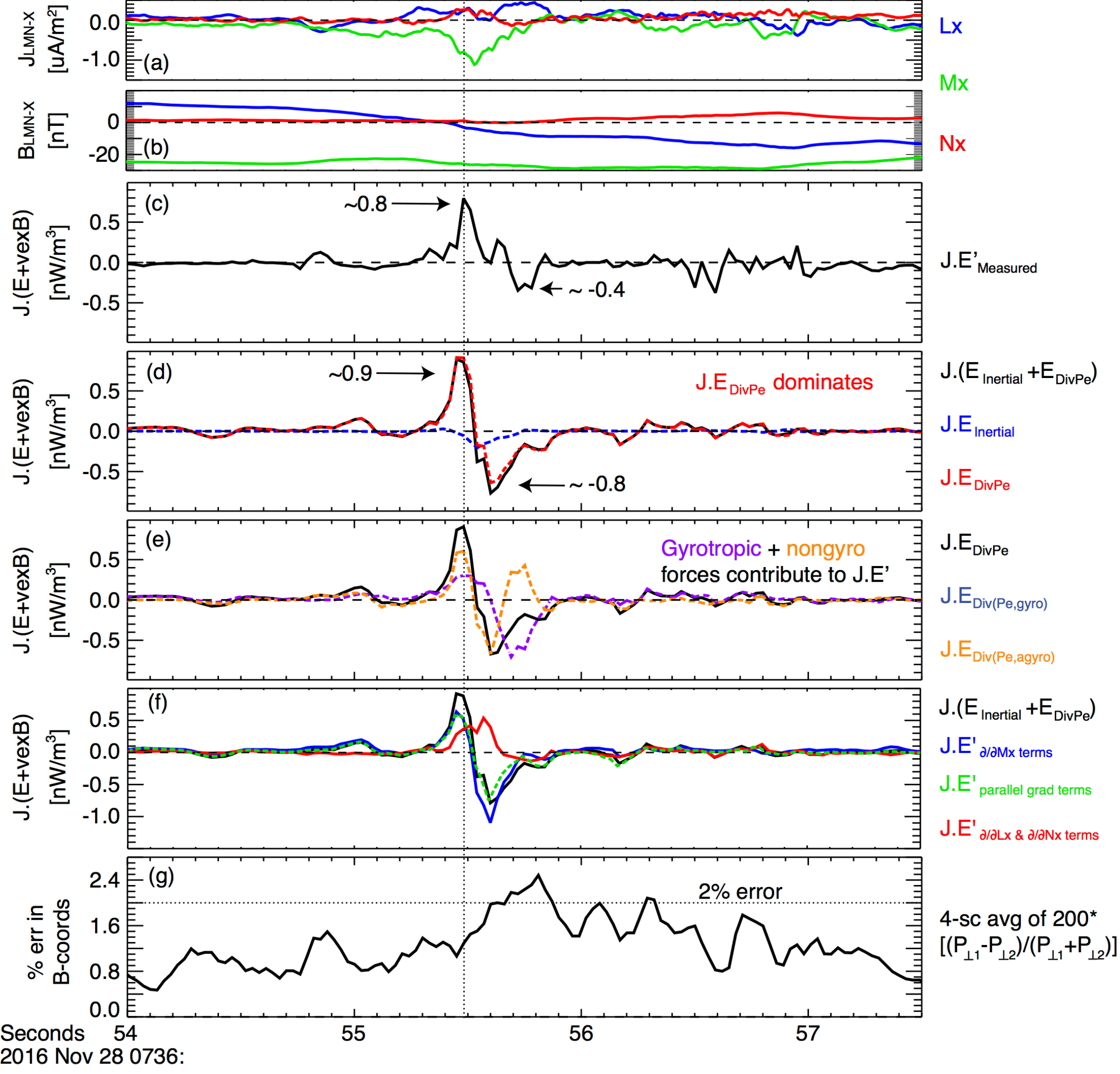}
\caption{Analysis of generalized Ohm's law near the $B_L$ reversal point, as shown in (a), and the intense out-of-plane electron current layer, as shown in (b). (c)--(f) the non-ideal energy conversion rate determined by (c) the measured electric field and (d)--(f) the electric field approximated from the plasma electron data as $\vec{E}'\approx\vec{E}_{DivPe}+\vec{E}_{inertial}$. (d) A comparison of the two terms in the approximated $\vec{J}\cdot\vec{E}'$. (e) A comparison of the gyrotropic versus agyrotropic contributions to $\vec{J}\cdot(\vec{E}_{DivPe}+\vec{E}_{inertial})$. (f) A comparison of the in-plane versus out-of-plane gradient terms in the pressure divergence. (g) A proxy for the error in the barycentric magnetic-pressure coordinates.}
\label{Ohms}
\end{figure*}

Figure \ref{Ohms} presents an analysis of the terms in generalized Ohm's law (equation 1) near the $B_L$ reversal region. For context, panels (a) and (b) show the magnetic field and current density vectors in $LMN-X$. Panel (c) shows $\vec{J}\cdot\vec{E}'$, where the measured electric field, electron velocity, and DC magnetic field vector have been used to calculate $\vec{E}'$ and the curlometer technique has been used to calculate the barycentric $\vec{J}$. Panels (d)--(e) also show the average $\vec{J}\cdot\vec{E}'$, where the electron pressure tensor divergence $\vec{E}_{DivPe}=-\nabla\cdot\bar{P}_e/en$ and electron inertial $\vec{E}_{inertial}=-m_e\nabla\cdot(\vec{v}_e\vec{v}_e)/en$ terms have been used to approximate $\vec{E}'$. The contribution of the ion inertial term to $\vec{J}\cdot\vec{E}'$ is negligible. We have not accounted for the terms in Ohm's law that represent time-dependence of $\vec{J}$ and anomalous resistivity, which appear in equation (1). $\vec{E}_{DivPe}$ and $\vec{E}_{inertial}$ have been calculated with the standard 30-ms resolution electron data. We have calculated the same terms with the 7.5-ms resolution data of \textit{Rager et al.,} [submitted], and we found no considerable differences in the form of either $\vec{E}_{DivPe}$ or $\vec{E}_{inertial}$, which suggests that the electron distributions were well resolved at 30-ms.

First, we compare the measured $\vec{J}\cdot\vec{E}'$ with the approximated $\vec{J}\cdot(\vec{E}_{DivPe}+\vec{E}_{inertial})$, given by the black curves in Figure \ref{Ohms}c and d, respectively. The measured energy conversion rate peaks at $\sim$0.8 nW/m$^3$ at 7:36:55.5. This positive peak is followed by a smaller negative peak of $\sim$--0.4 nW/m$^3$ at 7:36:55.7--55.8. Both the magnitude and the timing of the positive peak are approximated very well by $\vec{J}\cdot(\vec{E}_{DivPe}+\vec{E}_{inertial})$. The magnitude and timing of the negative peak are misestimated, being too early by 0.15 seconds and too large by a factor of $\sim$2. These errors are likely a result of our approximation of generalized Ohm's law by two of four terms and/or our approximation of gradients as linear. Still, we conclude that the approximation of $\vec{J}\cdot\vec{E}'\approx\vec{J}\cdot(\vec{E}_{DivPe}+\vec{E}_{inertial})$ does a good job at describing the quality of the curve and a fair job at describing its quantity. The oscillations in the measured $\vec{J}\cdot\vec{E}'$, which are seen in Figure \ref{Ohms}c between roughly 56.2--57 seconds, correspond to the region highlighted in Figure \ref{4probe_plot}e--g as the magnetosheath-side separatrix. There are no obviously correlated oscillations in our approximation of $\vec{J}\cdot\vec{E}'$ with these two terms of generalized Ohm's law.

The red and blue dashed lines on Figure \ref{Ohms}d represent the portions of $\vec{J}\cdot\vec{E}'$ driven by electron pressure divergence and electron inertia, respectively. As is evident, the electron pressure divergence term completely dominates the energy conversion rate and nearly independently defines both the maximum and minimum of the total (black). The largest value of $\vec{J}\cdot\vec{E}_{inertial}$ is $\sim$-0.2 nW/m$^3$, which, in an absolute sense, is roughly $\sim$20\% the maximum value of $\vec{J}\cdot\vec{E}_{DivPe}$. The inertial electric field is also at least partially anti-aligned with $\vec{J}$, as is evident by the negative value of their inner product. Unlike the pressure divergence term, the largest value of $\vec{J}\cdot\vec{E}_{inertial}$ is almost exactly aligned in time with the largest out-of-plane current. The largest values of both curves, though, are achieved within the out-of-plane current layer.

The black curve in Figure \ref{Ohms}e, $\vec{J}\cdot\vec{E}_{DivPe}$, is broken up into two component parts; the magenta curve represents the divergence of the gyrotropic portion of the pressure tensor and the orange curve represents the divergence of the agyrotropic portion. As is evident, both terms seem to play some role in governing the energy conversion rate near the X-point. At the exact X-point, which we have suggested to have been northward of MMS, it is expected that the agyrotropic pressure force dominates the gyrotropic force completely, whereas the opposite is expected outside the EDR [\textit{Hesse et al.,} submitted]. In the strong positive peak of $\vec{J}\cdot\vec{E}'$, the gyrotropic term is roughly half as large as the agyrotropic term. Here, both terms are positive. The story is more complicated in the negative peak of $\vec{J}\cdot\vec{E}'$, where both terms have both large positive values and large negative values in such a way that the two are partially balanced. The two terms are not fully balanced, however, and the intersection between the two curves coincides with the location of the negative peak of the overall $\vec{J}\cdot\vec{E}_{DivPe}$ curve. We interpret these results, in a general sense, as an indication that MMS was located within the EDR, where agyrotropic pressure forces are non-negligable, but outside the central EDR, where gyrotropic forces are negligible. 

In order to calculate $\nabla\cdot\bar{P}_{e,gyro}$ and $\nabla\cdot\bar{P}_{e,agyro}$, we have organized the pressure tensor with magnetic field coordinates local to the spacecraft tetrahedron. The parallel axis is defined by the four-spacecraft-averaged field vector. The two perpendicular axes are defined such that the last two diagonal elements of the four-spacecraft-averaged pressure tensor are equivalent, i.e., $\left<P_{e,\bot1}(t)\right>=\left<P_{e,\bot2}(t)\right>$, where the brackets indicate averaging over the four spacecraft and the coordinate system is time dependent. In this coordinate system, the gyrotropic elements are the diagonals and the agyrotropic elements are the off-diagonals. Invariably, given the finite separation of the four spacecraft, this coordinate system will not separate the gyrotropic and agyrotropic pressures exactly and simultaneously for all four spacecraft. By means of estimating the error in this technique, we have calculated the four-spacecraft-averaged difference between the two perpendicular diagonal pressures, $\left<2(P_{e,\bot1}-P_{e,\bot2})/(P_{e,\bot1}+P_{e,\bot2})\right>$, which is shown in Figure \ref{Ohms}g. The average error is on the order of roughly 2\% and the maximum error for a single spacecraft is less than 4\% (not pictured). We interpret this as an indication that the gyrotropic and agyrotropic portions of the pressure tensor can be separated simultaneously on all four spacecraft with sufficient accuracy to warrant the qualitative interpretation offered in the previous paragraph.

Finally, we separate the divergences in $\vec{J}\cdot(\vec{E}_{DivPe}+\vec{E}_{inertial})$ by the directions of the derivatives. In Figure \ref{Ohms}f, $\nabla_{in-plane}\cdot\bar{P}_{e}\equiv(\hat{L}\partial/\partial L)\cdot\bar{P}_{e}+(\hat{N}\partial/\partial N)\cdot\bar{P}_{e}$ is given by the red curve and $\nabla_{out-of-plane}\cdot\bar{P}_{e}\equiv(\hat{M}\partial/\partial M)\cdot\bar{P}_{e}$ is given by the blue curve. This separation by directional derivative is done for the combination of the much larger pressure divergence term and the much smaller inertial term. Since this sort of analysis may suffer if, for instance, $N$ and $M$ are not separated precisely, we complement the out-of-plane $\partial/\partial M$ term by calculating the gradients along the magnetic field direction (dashed/green curve in \ref{Ohms}f). At or very near the X-point of guide-field reconnection where $B_L \approx B_N\approx0$, the out-of-plane direction is the direction of the magnetic field, $B_M$. We expect this approximation to work very well near the large positive $\vec{J}\cdot\vec{E}'$ peak since it is nearest the $B_L$ reversal, and not very well for the large negative $\vec{J}\cdot\vec{E}'$ peak as it is sheath-ward of the $B_L$ reversal.

Indeed, both methods for estimating the out-of-plane gradients yield similar curves near the positive peak of $\vec{J}\cdot\vec{E}'$, then diverge during the negative peak. We interpret this as an indication that we have separated the in and out-of-plane gradients with sufficient accuracy to claim that roughly 50\% of the field-to-plasma energy conversion in this EDR was being driven by structures in the out-of-plane direction. The portion of $\vec{J}\cdot(\vec{E}_{DivPe}+\vec{E}_{inertial})$ from in-plane gradients is weak but entirely positive, which is expected in the central EDR \cite{Zenitani.2011,Shay.2016}, whereas the portion coming from out-of-plane gradients seems to be solely responsible for the negative energy conversion rate.

\section{Conclusions}

\subsection{Summary} 
 
This study identified and analyzed an electron diffusion region (EDR) event observed at the dayside magnetopause by Magnetospheric Multiscale (MMS) on 2016-11-28 at 7:36 UT. The magnetopause was characterized by a moderate magnetic shear angle of $\sim129^\circ$, a large density asymmetry of $n_{sh}/n_{sp}\approx27$, and a large asymmetry in the reconnecting magnetic field component of $B_{L,sh}/B_{L,sp}\approx0.4$. Additional upstream parameters are in Table 1. The EDR was observed at a time when the MMS tetrahedron was extremely small (6.4 km), which permitted the analysis of spatial gradients of the plasma moments and field data.

Finite-gyro-penetrating magnetosheath protons were observed near but magnetosphere-ward of the electron separatrix, indicating that the gyro-radius reconnecting layer was sub-ion-scale at the location of MMS at the start of the crossing. Similar signatures of finite-gyro-penetrating magnetospheric protons were also observed, as in \cite{Phan.2016them}. Electron ``outflow'' crescents were observed within the magnetosheath-ward side of the magnetospheric separatrix, which are remote signatures of the mixing of inflowing plasmas in the central EDR between the stagnation and X points \cite{Shay.2016,Hwang.2017}. Filamentary electron velocities similar to those of \cite{Phan.2016} and \cite{WangRS.2017} were observed during the crossings of the magnetosphere-side separatrix. Electromagnetic whistler waves were observed prior to the final crossing, when MMS was near/in the magnetosphere-side inflow region, as predicted by \cite{Fujimoto.2014}. Low-frequency electrostatic waves similar to those of \cite{Ergun.2016b,Ergun.2017} were observed at both the electron separatrix ($\pm L$-directed flow layer) and out-of-plane electron current layer. Following the crossing of the magnetospheric separatrix, a strong Hall $B_M<0$ was observed. The Hall field remained large and negative throughout the crossing, indicating that MMS did not cross or enter the central electron diffusion region. No super-Alfvenic perpendicular electron jets were observed, despite evidence for ion demagnetization.

Near the X-point, at the reconnection site mid-plane, all four MMS nearly simultaneously encountered signatures of the EDR, including an intense and thin electron current layer, electron agyrotropy and anisotropy, non-ideal electric fields, non-ideal energy conversion, etc. We understood the mis-matched timing of the regions where $\lambda\leq \rho_{e}$ and the electron agyrotropy was significant to be an indication that the LMN coordinates found for the whole crossing did not organize the data near the X-point. We then found a new LMN-X system where the inequality of \cite{Hesse.2014,Hesse.2016} ($\lambda\leq \rho_{e}$) was satisfied during the period with the largest electron agyrotropy. 

The electrons surrounding the current layer had a roughly symmetric flat-top-type distribution in $\pm\vec{v}_{\parallel}$. The symmetry of the flat-top was broken in the electron current layer, where the distribution was significantly more extended in the anti-parallel direction. The streaming of the electrons in the anti-parallel direction was responsible for the current and likely related to the colocated $\vec{J}\cdot\vec{E}'>0$. The $\vec{v}_{\parallel}-\vec{v}_{\bot}$ distribution functions at the X-point were similar in character to those observed during large guide field reconnection by \cite{Eriksson.2016} and during moderate guide field reconnection by \cite{Genestreti.2017}.

The directly-measured energy conversion rate at the X-point was reasonably well approximated by $\vec{J}\cdot(\vec{E}_{DivPe}+\vec{E}_{inertial})$. The pressure divergence term dominated the inertial term by a wide margin. This is similar to MMS observations of the stagnation point during high-shear reconnection \cite{Torbert.2016b} [\textit{Rager et al.} submitted], but dissimilar to the predictions of 2.5-d PIC simulations \cite{Hesse.2014,Hesse.2016}. Further analysis of the pressure divergence term revealed that both the gyrotropic and agyrotropic pressure forces contributed to the overall energy conversion rate, but the meaning behind the structure and specific balance of the two terms is not currently clear. Finally, by separating the directional derivatives in the two Ohm's law terms, we found that both out-of-plane and in-plane gradients contributed to the inertial and pressure divergence terms. The portion of $\vec{J}\cdot(\vec{E}_{DivPe}+\vec{E}_{inertial})$ due to in-plane gradients was smaller and positive, whereas the portion from out-of-plane gradients was almost entirely responsible for the negative peak of $\vec{J}\cdot\vec{E}'$.

\subsection{Discussion and future work}

The picture of this single EDR, which was obtained by the unique capabilities of the MMS suite of plasma instruments, differs slightly from the picture of laminar and steady-state 2.5-d PIC simulations. Both (a) the importance of electron agyrotropy during $\vec{J}\cdot\vec{E}'>0$ at the X-point and (b) the observation of electron outflow crescents indicate the importance of kinetic-scale mixing of inflowing plasmas in the central EDR, which is in line with previous MMS observations \cite{Burch.2016b}, 2.5-d PIC simulations \cite{Hesse.2014,Hesse.2016,Shay.2016}, and 3-d PIC simulations \cite{Price.2016,Price.2017}. The dominance of the electron pressure divergence term differs from the predictions of 2.5-d simulations, as does the considerable contribution from out-of-plane gradients of the pressure tensor.

\begin{figure*}
\noindent\includegraphics[width=39pc]{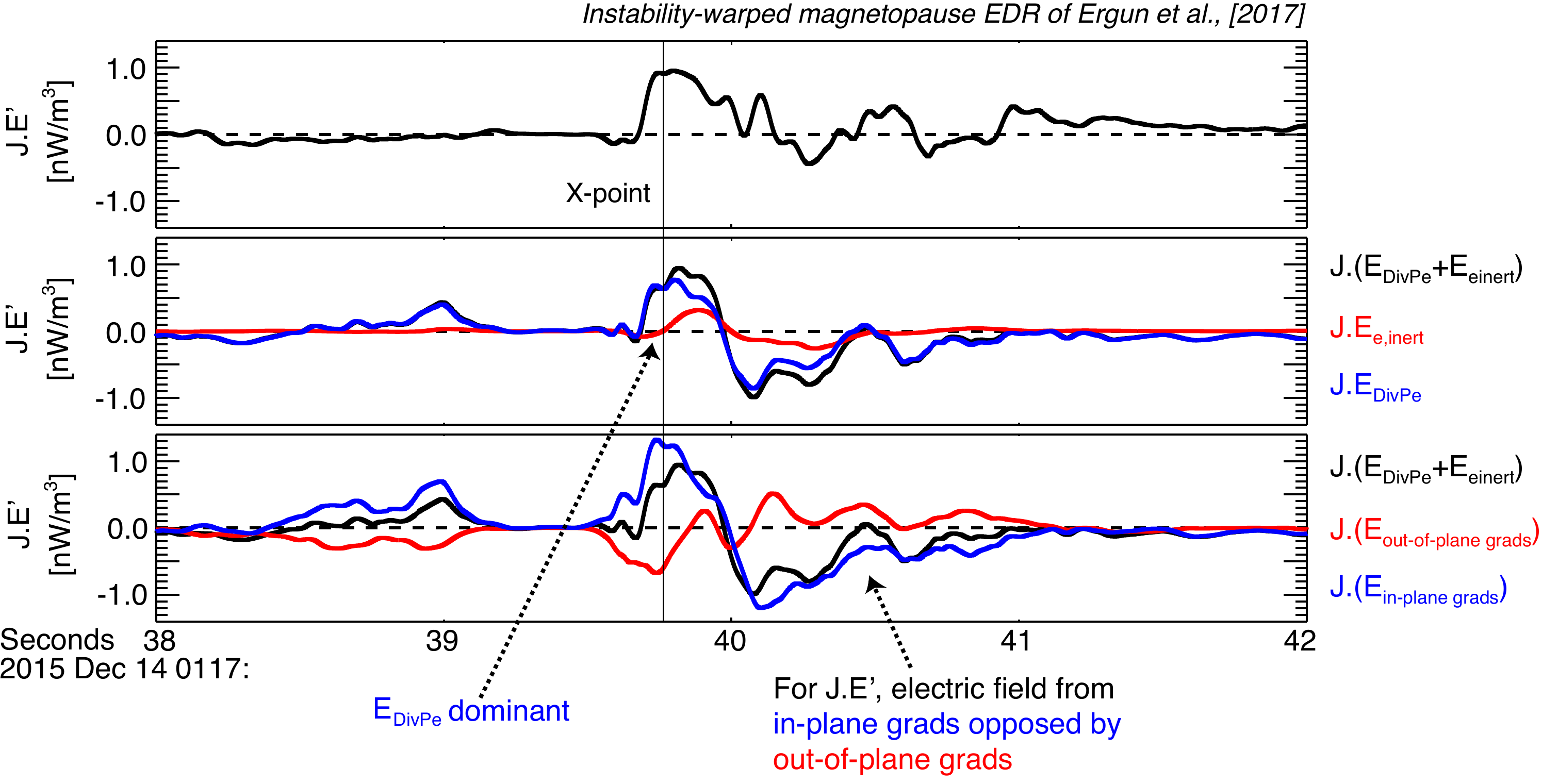}
\caption{Analysis of generalized Ohm's law, (a) similar to Figure \ref{Ohms}c, (b) similar to \ref{Ohms}d, (c) similar to \ref{Ohms}f, but for the corrugated magnetopause event of \cite{Ergun.2017}.}
\label{ergunplot}
\end{figure*}

We suggest that the $\hat{M}{(\partial/\partial M)} \bar{P}_e$ term is caused by the wrapping of the $\hat{N}{(\partial/\partial N)} \bar{P}_e$ term into the $M$ direction, as was predicted by \cite{Price.2017}. This would explain (1) the origin of the out-of-plane gradient terms, and (2) why the pressure tensor divergence contributed dominantly to $\vec{J}\cdot\vec{E}'$ near our X-point, seeing as $\hat{N}{(\partial/\partial N)} \bar{P}_e$ is expected to be nearly an order of magnitude larger than $E_M$ in 2.5-D simulations \cite{Shay.2016}. The out-of-plane current varies significantly in the out-of-plane direction ($(J_M/d_{e,asym})^{-1}\cdot\partial J_{M}/\partial M\sim15\%$), which is qualitatively consistent with the picture of \citeauthor{Price.2017}. More evidence is likely needed to support the relevance of this scenario to this particular event, including some analysis of the 3-d structure of the intense out-of-plane current layer observed near the $B_L$ reversal. We have, however, found similar results for one of the corrugated magnetopause events of \cite{Ergun.2017}, as is shown in Figure \ref{ergunplot}. Energy conversion near the X-point is driven predominantly by the pressure divergence term, which is approximately twice as large as the inertial term. Similar to the 2016-11-28 event, the terms resulting from out-of-plane gradients are considerable, though this time 50\% as large as the in-plane gradient terms and at least partially anti-aligned with the current.

In the future, it would also be important to analyze 3-d simulations with similar upstream conditions to those provided in Table 1. This could be done in order to determine if $\vec{J}\cdot\vec{E}'$ and the terms in generalized Ohm's law should vary along the corrugations as $\vec{J}$ is thought to \cite{Price.2016}. Finally, it is also desirable to analyze other EDR events in the same way that we have analyzed the two events in Figures \ref{Ohms} and \ref{ergunplot}. While we analyzed additional events from \cite{Ergun.2017}, none had the same clear correlation between the measured and approximated forms of $\vec{J}\cdot\vec{E}'$, which may indicate that the linear gradient assumption was invalid for these events.



%
%
%
%
%
%
%

\begin{acknowledgments}
The authors would like to thank everyone who contributed to the success of the MMS mission and those who contributed to the rich scientific heritage on which this mission is based. KJG was funded by the FFG project number 847969. RN was supported by Austrian Science Funds (FWF) I2016-N20. ZV was supported by FWF grant P28764-N27. JPE was supported by the United Kingdom Science Technology Facilities Council (STFC) under grant ST/N000692/1. This event was identified during an International Space Science Institute (ISSI) meeting of the``MMS and Cluster observations of magnetic reconnection'' group. KJG would like to thank the EDP team, specifically Dr. Narges Ahmadi, for providing the l3 electric field data. KJG would also like to thank Dr.s Amy Rager, John Dorelli, and Daniel Gershman for providing us with the 7.5-ms electron data. Excepting these two data products, which are available upon request, the MMS data was obtained from the MMS Science Data Center (SDC) at the University of Colorado. The study made use of the Space Physics Environment Data Analysis Software (SPEDAS) package.
\end{acknowledgments}

\end{article}
%
%
%
%
%
%
%
%


\end{document}